# OBSTRUCTIONS OF TURKISH PUBLIC ORGANIZATIONS GETTING ISO/IEC 27001 CERTIFIED

Tolga MATARACIOGLU[1], Sevgi OZKAN YILDIRIM[2]
[1]TUBITAK BILGEM Cyber Security Institute, Turkey
[2]Middle East Technical University, Informatics Institute, Dept. of Information Systems, Turkey

*ABSTRACT*

*In this paper; a comparison has been made among the Articles contained in the ISO/IEC 27001 Standard and the Articles of the Civil Servants Law No 657, which should essentially be complied with by the personnel employed within the bodies of public institutions in Turkey; and efforts have been made in order to emphasize the consistent Articles; and in addition, the matters, which should be paid attention by the public institutions indenting to obtain the ISO/IEC 27001 certificate for the Articles of the Civil Servants Law No 657 which are not consistent with the ISO/IEC 27001 certification process, have been mentioned. Furthermore, solution offers have been presented in order to ensure that the mentioned Articles become consistent with the ISO/IEC 27001 certification process.*

*KEYWORDS*

*Information Security Management System, the Civil Servants Law No 657, the ISO/IEC 27001 Standard*

*ABBREVIATIONS*

| BILGEM: | Center of Research for Advanced Technologies of Informatics and Information Security |
|---|---|
| IEC: | International Electrotechnical Commission |
| ISMS: | Information Security Management System |
| ISO: | International Organization for Standardization |
| TUBITAK: | The Scientific and Technological Research Council of Turkey |

## 1. INTRODUCTION

Information security is defined as the operation indented for protection of information against a large cluster of threats in order to ensure business continuity, in order to be able to minimize business damages/losses and in order to increase the yields/returns of investments and business opportunities [1]. Today, security vulnerabilities occur and, those security vulnerabilities may arise from further inclusion of information technologies into daily life, from use of information technologies by a larger number of persons, from production of information technology products expeditiously and, from development of systems rapidly [1].

Fundamental components of information security are confidentiality, integrity and availability [2][3][4]. Confidentiality refers to the circumstances where it is guaranteed that only authorized persons are entitled to access information. Cryptographic measures, access control and firewalls

International Journal of Managing Value and Supply Chains (IJMVSC) Vol.5, No. 2, June 2014

are the leading precautions that ensure protection of confidentiality. Integrity ensures that information and communication systems are allowed to be modified only by authorized persons and operations. Hashing, antivirus systems and patch management are the leading measures that ensure protection of integrity. Availability is defined as the guarantee whereby authorized persons access information at any time that they desire and within the period of time in which they are authorized. Backing up is the primary measure among the other measures that ensure protection of availability [5][6][7].

A framework, which designs, authenticates, operates, monitors, assesses, maintains and develops information security in an institution, is provided by ISMS [2]. ISMS lends assistance in attainment of the following articles [8] [9][10][11]:

•	To be able to manage the risks that occur in business activities,
•	To be able to manage the activities aimed at responding to security incidents,
•	To ensure that a security culture is constituted within the body of the institution.

Today, in addition to the private sector, it has become almost a requirement to establish ISMS and obtain ISO/IEC 27001 certificate also in public institutions as a consequence of the inspections conducted particularly by the Supreme Court of Public Accounts. The Articles, which should essentially be complied with by the personnel employed within the bodies of public institutions in Turkey, are listed in the Civil Servants Law No 657; and, the rules, which the institutions intending to establish ISMS and obtain ISO/IEC 27001 certificate should comply with, are listed in the ISO/IEC 27001 Standard [2][3][12].

The Civil Servants Law No 657 has been adopted on July 14th, 1965 and has been published in the Official Gazette No 12056 and dated July 23rd, 1965, and various Articles of the Law have been amended within the period between the years 1970 and 2011 [12].

In terms of ISMS establishment process; even though, in respect of public institutions, there is mostly a consistency among the Articles of the Law No 657 and the Articles contained in the ISO/IEC 27001 Standard; it is not possible to say that there is such a consistency in terms of certain Articles contained in the Law No 657. The inconsistencies, which are described in this paper, have been encountered in many public institutions with which TUBITAK BILGEM Cyber Security Institute provides consultancy services relevant to ISMS establishment and ISO/IEC 27001 certification; and therefore, this fact has constituted the most significant factor for issuance of this paper.

Annex A.15.1 of the ISO/IEC 27001 Standard contains the controls relevant to identification of applicable legislation. In Annex A.15.1.1, it is stated that "All relevant statutory, regulatory and contractual requirements and the organizations approach to meet these requirements shall be explicitly defined, documented, and kept up to date for each information system and the organization" [2].

During determination of the public institutions' legal obligations within the ISMS establishment process; in addition to the Law No 657, also the Turkish Criminal Code, the Law No 5651 on Regulating Broadcasting in the Internet and Fighting Against Crimes Committed through Internet Broadcasting, the Law No 5846 on Intellectual and Artistic Works, the Regulation on Principles of Ethical Conduct for Public Officials with the Regulation on Application Procedures and Principles, the Regulation on the Procedures and Principles for Regulating Broadcasting in the Internet Environment, the Regulation on Collected Usage Internet Providers, the Regulation concerning State Archives Services and, the Prime Ministry Circular on Use of Licensed Software should be reviewed.

In this paper; a comparison has been made among the Articles contained in the ISO/IEC 27001 Standard and the Articles of the Civil Servants Law No 657, which should essentially be complied with by the personnel employed within the bodies of public institutions in Turkey; and efforts have been made in order to emphasize the consistent Articles; and in addition, the matters, which should be paid attention by the public institutions indenting to obtain the ISO/IEC 27001 certificate for the Articles of the Civil Servants Law No 657 which are not consistent with the ISO/IEC 27001 certification process, have been mentioned. Furthermore, solution offers have been presented in order to ensure that the mentioned Articles become consistent with the ISO/IEC 27001 certification process.

## II. METHOD

The Articles contained in the Civil Servants Law No 657 and the Articles contained in the ISO/IEC 27001 Standard have been reviewed and; after the Articles, which are relevant to information security, have been selected from within the Articles contained in the Civil Servants Law No 657, the equivalent Articles contained in the ISO/IEC 27001 Standard have been found and thus, a mapping has been performed among the Articles of 657&27001. During the mapping, in addition to the mandatory Articles in the ISO/IEC 27001 Standard, also the 133 controls, which are contained in the Annex A, have been dealt with. Thereafter, from within the Articles contained in the Civil Servants Law No 657, the Articles consistent with the ISO/IEC 27001 Standard and the Articles, which are not consistent with the ISO/IEC 27001 certification process, have been separated. In the light of the problems, which are, in practice, observed in the public institutions with which consultancy services are provided in relation to ISMS establishment and ISO/IEC 27001 certification; the Articles of the Civil Servants Law, which are not consistent with the ISO/IEC 27001 certification process, have been discovered.

## III. RESULTS

In this section, the Articles of the Civil Servants Law No 657, which are consistent with the ISO/IEC 27001 Standard, and, the Articles of the Civil Servants Law No 657, which are not consistent with the ISO/IEC 27001 Standard, have been analyzed and listed.

In Table 1; the Articles of the Civil Servants Law No 657, which are consistent with the ISO/IEC 27001 Standard, have been listed so that they are mapped to the mandatory articles of the standard or controls listed in Annex A of the ISO/IEC 27001 Standard. The Articles contained in the Table 1 do not represent a "full list" and thus, Table 1 includes only the most critical Articles. In addition to those Articles, there are many Articles in the Law No 657, which are consistent with the ISO/IEC 27001.

Table 1: The Articles of the Civil Servants Law No 657, which are consistent with the ISO/IEC 27001 Standard [2][12]

| No | The Article of the Civil Servants Law No 657 and the Explanation Thereof | The Equivalent under the ISO/IEC 27001 Standard |
|----|--------------------------------------------------------------------------|------------------------------------------------|
| 1  | Article 3 A "Classification": is to classify the public services and the civil servants, who are employed for those services, in accordance with the characteristics and qualifications required by those services and in accordance with the occupations/professions. | A.6.1.3 Allocation of information security responsibilities: All information security responsibilities shall be clearly defined.<br><br>A.8.1.1 Roles and responsibilities: Security roles and responsibilities of employees, contractors and third party users shall be defined and documented in accordance with the organizations |

| No | The Article of the Civil Servants Law No 657 and the Explanation Thereof | The Equivalent under the ISO/IEC 27001 Standard |
|---|---|---|
|  |  | information security policy. |
| 2 | Article 7: Civil servants are, in all cases, obligated to protect benefits of the State. They shall not perform any activities which are contrary to the Constitution and the Laws of the Republic of Turkey, which prejudice the integrity and independence of the Country and which endanger the security of the Republic of Turkey. | A.6.1.5 Confidentiality agreements: Requirements for confidentiality or non-disclosure agreements reflecting the organizations needs for the protection of information shall be identified and regularly reviewed.<br><br>A.15.1.4 Data protection and privacy of personal information: Data protection and privacy shall be ensured as required in relevant legislation, regulations, and, if applicable, contractual clauses. |
| 3 | Article 10: In the units of institutions and services, of which civil servants are the chief; those civil servants are entrusted with and responsible for timely and completely fulfillment of the services determined in laws, by-laws and regulations as well as having those services fulfilled timely and completely and, are entrusted with and responsible for training their civil servants employed as their subordinates and, are entrusted with and responsible for monitoring and inspecting their acts and behaviors. | 5.2.2 Training, awareness and competence: The organization shall ensure that all personnel who are assigned responsibilities defined in the ISMS are competent to perform the required tasks by:<br><br>a) determining the necessary competencies for personnel performing work effecting the ISMS;<br>b) providing training or taking other actions (e.g. employing competent personnel) to satisfy these needs;<br>c) evaluating the effectiveness of the actions taken; and<br>d) maintaining records of education, training, skills, experience and qualifications.<br><br>The organization shall also ensure that all relevant personnel are aware of the relevance and importance of their information security activities and how they contribute to the achievement of the ISMS objectives. |
| 4 | Article 12: Civil Servants are obligated to fulfill their duties carefully and meticulously and, are obligated to protect the Government properties delivered to them and, are obligated to take required measures in order to keep them available for service at any time. | A.9.2 Equipment security: To prevent loss, damage, theft or compromise of assets and interruption to the organizations activities. |
| 5 | Article 16: Civil Servants are not entitled to take any official documents, tools and instruments, which are relevant to their duties, away from the places for which the power has been granted and, are not entitled to use them for their private matters and affairs. Civil Servants are obligated to return official documents, tools and instruments, which have been delivered to them as required by their duties, at the time when their commissions expire or are terminated. This obligation covers also inheritors of civil servants. | A.9.2.7 Removal of property: Equipment, information or software shall not be taken off-site without prior authorization.<br><br>A.8.3.2 Return of assets: All employees, contractors and third party users shall return all of the organizations assets in their possession upon termination of their employment, contract or agreement. |

International Journal of Managing Value and Supply Chains (IJMVSC) Vol.5, No. 2, June 2014

| No | *The Article of the Civil Servants Law No 657 and the Explanation Thereof* | *The Equivalent under the ISO/IEC 27001 Standard* |
|---|---|---|
| 6 | *Article 31: Civil Servants are not allowed to disclose any confidential information relevant to the public services, without the written permission of the authorized Minister, even if they have resigned or retired from office.* | *A.6.1.5 Confidentiality agreements: Requirements for confidentiality or non-disclosure agreements reflecting the organizations needs for the protection of information shall be identified and regularly reviewed.*<br><br>*A.15.1.4 Data protection and privacy of personal information: Data protection and privacy shall be ensured as required in relevant legislation, regulations, and, if applicable, contractual clauses.* |
| 7 | *Article 125 (only a part thereof has been dealt with): The disciplinary punishments to be inflicted on Civil Servants, and the acts and behaviors, which require disciplinary punishments, are as follows:*<br><br>*B - Reprimand: is to notify in written form the civil servant that he/she is faulty in his/her duty and behaviors. The acts and behaviors, which require reprimand, are as follows:*<br><br>*a) Acting faultily in completely and timely fulfillment of the instructions given and of the duties assigned, acting faultily in fulfillment of the principles and procedures determined in his/her place of duty by the relevant institution and, acting faultily in protection, use and maintenance of any official documents, tools and instruments relevant to his/her duty,*<br><br>*e) Use of any official documents, tools and instruments, which belong to the State, by him/her for his/her private matters and affairs,*<br><br>*f) Losing any official documents, tools, instruments and similar properties belonging to the State*<br><br>*D – Suspension of Promotion: is to suspend promotion of the civil servant for a period of time between 1 year and 3 years in his/her current grade, depending on the severity level of the relevant act. The acts and behaviors, which require the suspension of promotion, are as follows:*<br><br>*k) Disclosure of any information, the disclosure of which is prohibited* | *A.7.1.3 Acceptable use of assets: Rules for the acceptable use of information and assets associated with information processing facilities shall be identified, documented, and implemented.*<br><br>*A.8.2.3 Disciplinary process: There shall be a formal disciplinary process for employees who have committed a security breach.*<br><br>*A.15.1.4 Data protection and privacy of personal information: Data protection and privacy shall be ensured as required in relevant legislation, regulations, and, if applicable, contractual clauses.* |

We list the Articles of the Law No 657, which are not consistent with the ISMS establishment and ISO/IEC 27001 certification process in Table 2. Those listed inconsistencies have been determined in the light of the problems observed in the public institutions with which consultancy services are provided in relation to ISMS establishment and ISO/IEC 27001 certification.

Table 2: The Articles of the Law No 657, which are not consistent with the ISO/IEC 27001 certification process [12]

| No. | *The Article 657 of the Civil Servants Law No 657 and the Explanation Thereof* |
|---|---|
| 1 | *Article 8: It is a principle that civil servants work in cooperation with each other.* |
| 2 | *Article 12: If the administration suffers a loss/damage due to a willful, faulty, negligent or incautious act of a civil servant; it is a principle that such loss/damage is indemnified by this civil servant on the basis of current market value of that loss/damage.* |
| 3 | *Article 15: Civil Servants are, in relation to civil services, not entitled to disclose any information and make any statements to the press, news agencies or radio and television corporations. Required information relevant those matters can be furnished only by the governors in the commissioned provinces to be empowered by the Minister or by the official to be empowered by the governors.* |
| 4 | *Article 125: B - Reprimand: is to notify in written form the civil servant that he/she is faulty in his/her duty and behaviors. The acts and behaviors, which require reprimand, are as follows:*<br><br>*j) Raising objections against the instructions given*<br><br>*m) Disclosure of information or making statements to the press, news agencies or radio and television corporations, though he/she is not authorized* |

### i. Table 2 - No.1

According to the Article 4.2.1.a of the ISO/IEC 27001 Standard; the Institution should define the scope and boundaries of the ISMS in terms of the characteristics of the business, the organization, its location, assets, technology, and including details of and justification for any exclusions from the scope. In general, public institutions tend to prefer a narrow scope in terms of manageability. The personnel and physical boundaries that are included in the preferred scope represent the institution in terms of ISMS. However, as required by the ISMS establishment, it is necessary for the institution to act in cooperation with other departments. If, resulting from a risk analysis, purchase of a hardware is required, it is necessary to come into contact with the Department of Administrative and Financial Affairs; if preparation of a confidentiality agreement is required, it is necessary to come into contact with the Legal Consultancy Department; if issuance of an end-of-service form is required, it is necessary to come into contact with the Human Resources Department. In practice, it is observed that the institutions, which have been determined within the scope, are not able to act in cooperation with other departments and tend to resolve problems at all times within their own bodies. Therefore, such a case constitutes a conflict with the Article 8 of the Civil Servants Law No 657.

### ii. Table 2 - No.2

According to the Article 4.2.1.e of the ISO/IEC 27001 Standard, it is necessary to perform an analysis and evaluate risks found in the institution. Within this scope, the values of the assets are determined. During determination of cumulative values of the assets, following method is used: The value of the campus is inherited from the institution building, the value of the institution building is inherited from the system center room, the value of the system center room is inherited from the equipments, the value of the equipments is inherited from the hardware, the value of the hardware is inherited from the software contained on them, and finally, the value of the software is inherited from the information that they run. In brief, when the value of information is determined, it means that the cumulative value of all the assets in the institution has been determined [13]. For example, the real value of a server situated in the system center room is related to the criticality of the information running on the server. In case a server becomes inoperative due to negligence of a civil servant; the civil servant shall pay only the amount corresponding to the current market value of the server, as required by the Article 12 of the Civil Servants Law No 657. However, it is not possible for the institution to calculate the monetary value of the information lost on the server. Therefore, according to the Article 12, indemnification by the civil servant in relation to the damage arising from loss of that information will be skipped; and in such a case, the indemnification will remain insufficient for recovery of the entire loss suffered by the institution.

### iii. Table 2 - No.3

Annex A.14 of the ISO/IEC 27001 Standard indicates the method of business continuity. Within this scope, as required by Annex A.14.1.4, A single framework of business continuity plans shall be maintained to ensure all plans are consistent, to consistently address information security requirements, and to identify priorities for testing and maintenance. Determination of the personnel, who will fulfill the requirements of communications with third parties and the press in urgent cases, constitutes one of the sub-headings of the business continuity planning document to be prepared. As per the Article 15 of the Civil Servants Law No 657, this duty should be fulfilled only by the governors or by the official to be empowered by the governors. Particularly in the public institutions where a narrow scope is determined; it is not possible to make such an assignment.

### iv. Table 2 - No.4

According to Annex A.6.1.2 of the ISO/IEC 27001 Standard, which is relevant to coordination of information security; Information security activities shall be co-ordinated by representatives from different parts of the organization with relevant roles and job functions. Within this framework; at least an "Information Security Steering Committee" should be established within the body of the institution defined within the scope of ISMS. ISMS manager and the representatives of the work units included in the scope should constitute the members of this Committee. Within the body of the institution, there is hierarchy among the members of the Committee. Therefore, within the scope of the studies performed by the Information Security Steering Committee, if a committee member, whose position is superior than other committee members, has taken his/her final decision about a matter relevant to ISMS; subordinate members of the Committee are not entitled to raise any objections against such decision as per the Article 125 (j) of the Civil Servants Law No 657. Such a case plays a negative role in effectively performance of the ISMS studies. The explanations, which have been submitted for the Article 15 of the Civil Servants Law No 657, apply also to the Article 125 (m) thereof.

## IV. DISCUSSION

In this section, solution offers have been presented in order to ensure that the Articles of the Civil Servants Law No 657 contained in Table 2, which are not consistent with the ISO/IEC 27001 certification process, become consistent with the ISO/IEC 27001 certification process.

**i. Table 2 - No.1**

As to the matter that civil servants work in cooperation with each other within the scope of ISMS establishment, in other words, in ensuring cooperation among departments; the president of the public institution may, prior to ISMS establishment and by means of an internal circular, in other words, by means of an ISMS policy, announce all the employees of the institution that an ISMS will be established within the relevant scope and that, all employees and managers of the units should participate in those studies, if such a requirement arises.

As a second solution; an "Information Security Executive Committee" may be established within the body of the institution according to the information security coordination-related Annex A.6.1.2 of the ISO/IEC 27001 Standard. Regardless of how the scope of ISMS is determined; the president of the institution, and the chiefs of the legal consultancy, administrative and financial affairs and human resources departments as well as the chiefs of the departments included in the scope and ISMS manager may constitute the members of this Committee. Thus, any probable communication problems, which may be encountered among the departments, may be resolved radically.

As a third solution; it is possible to recommend spreading the scope of ISMS over the entire institution. In ensuring information security; it can be said that the ultimate point, which should be attained by the institution, is, in fact, determination of the scope as the whole institution.

**ii. Table 2 - No.2**

The most effective way to prevent the conflict encountered in terms of this Article may be to make an amendment on the Article 12 of the Civil Servants Law No 657. Current market value may be defined depending on only the own values of the tangible assets such as hardware, equipment and system center room. In addition, in the public institutions, a study may be conducted on how the cumulative value, which has been inherited by those assets from the information assets, contained on them, will be assessed and utilized in case those assets suffer a damage/loss. For example, in the risk analysis process, there may be a column in the assets inventory in relation to the tangible values of the assets; and meetings, which cover the entire institution and are intended to how the tangible value of the information will be determined, may be held in the institution.

**iii. Table 2 - No.3**

After the governors have granted the president of the institution the right to disclose information and make statements to the press, news agencies or to the radio and television corporations as per the Article 15 of the Civil Servants Law No 657; a definition may be provided in the business continuity plan (to be constituted within the body of the institution), which stipulates that, in urgent cases, the personnel, who will come into contact with third parties and the press, shall be the president of the institution. In such a case, in the level of institution presidency, participation in a part of the business continuity studies within the scope will be inevitable.

**iv. Table 2 - No.4**

As the most effective way to prevent the conflict encountered in terms of the Article 125 (j), in the direction of making an amendment on the relevant Article in the Civil Servants Law; it may be ensured that raising objections against the instructions given by the chiefs of the units do not result in a disciplinary punishment. Only under such an arrangement; the members of the Information Security Steering Committee are, in an equal and transparent environment, able to play an effective role in the process of discussing and resolving any ISMS problems.

The explanations, which have been submitted for the Article 15 of the Civil Servants Law No 657, apply also to the Article 125 (m) thereof.

## V. CONCLUSION

In conclusion; a comparison has been made among the Articles contained in the ISO/IEC 27001 Standard and the Articles contained in the Civil Servants Law No 657 and, in addition, efforts have been made in order to emphasize the consistent Articles; and furthermore, the matters, which should be paid attention by the public institutions indenting to obtain the ISO/IEC 27001 certificate for the Articles which are not consistent with the ISO/IEC 27001 certification process, have been mentioned in this paper. Moreover, solution offers have been presented in order to ensure that the mentioned Articles become consistent with the ISO/IEC 27001 certification process. Besides, solution offers have been presented for removal of the conflicts mentioned in Table 2. It can be said that, in fact, most of the Articles contained in the Law No 657 are consistent with the 27001 Standard and that, only for a few Articles listed in Table 2, public institutions should act carefully during the studies relevant to establishment of an ISMS. Within this scope, it is possible to utilize from the solution offers contained in Section IV.

**Tolga MATARACIOGLU**

After receiving his BSc degree in Electronics and Communications Engineering from Istanbul Technical University in 2002 with high honors, he received his MSc degree in Telecommunications Engineering from the same university in 2006. He is now pursuing his PhD degree in Information Systems from Middle East Technical University. He is working for TUBITAK BILGEM Cyber Security Institute as senior researcher. He is the author of many papers about information security published nationally and internationally. He also trains various organizations about information security. His areas of specialization are system design and security, operating systems security, and social engineering.

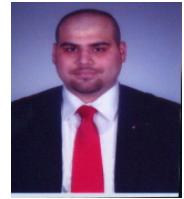

**Sevgi OZKAN YILDIRIM**

Dr Sevgi Ozkan Yildirim is an Associate Professor at Department of Information Systems, Informatics Institute, Middle East Technical University Turkey. She is currently the Associate Dean of the School. She holds a BA and an MA in Engineering from Cambridge University and an MSc in Business Information Systems London University UK. She has a PhD in Information Systems Evaluation. She is a Fellow of the UK Higher Education and a Research Fellow of Brunel University UK. Since 2006, Dr. Ozkan Yildirim has been involved with a number of EU 7th Framework and National projects in e-government.

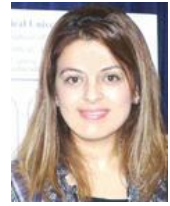